V. Mironov, J.P.M. Beijers

Kernfysisch Versneller Instituut, Zernikelaan 25, 9747 AA Groningen, The Netherlands

# THREE-DIMENSIONAL SIMULATIONS OF ION DYNAMICS IN THE PLASMA OF AN ELECTRON CYCLOTRON RESONANCE ION SOURCE.

*Abstract*

The ion production in an ECRIS is modelled using a particle-in-cell Monte-Carlo collision code in a three-dimensional geometry. Only the heavy particles (ions and atoms) are tracked, with the electron density determined from the requirement of quasi-neutrality, and the electron temperature is a free parameter. The electric fields in the plasma are assumed to be negligibly small, and the ion confinement due to a “potential dip” is neglected. It is found that experimentally observed features of ECRIS plasma are closely reproduced by the code, including the charge-state-distributions of extracted ion beams and sputtering patterns inside the source. The isotope anomaly is observed for the mixture of $^{20}Ne$ + $^{22}Ne$ isotopes, and some explanation for the effect is given. Possible connection between the wall-coating effect and parameters of the fast atoms created in collisions of the ions with the walls is discussed.

## INTRODUCTION

The growing demand for intense beams of highly charged ions stimulates further development of the Electron Cyclotron Resonance Ion Sources (ECRIS)[1]. This development is mainly done by following the semi-empirical scaling laws - increasing both the confining magnetic fields of the ECRIS trap and the frequency of the microwave radiation that heats electrons in the source plasma. Extracted ion currents can be increased by applying various special techniques[2] such as gas-mixing, wall-coating, afterglow and other effects. The physical mechanisms of these techniques are not yet clearly understood and are subject of discussion. No common agreement has been achieved so far on important aspects of ECRIS operation such as the mechanisms for the ion confinement in the plasma, electron energy distribution, shape of the plasma etc. Numerical simulations of ECR plasmas can pave the way for a better understanding of the physics of the source and assist in improving the source performance.

We report here on the development of a Particle-In-Cell Monte-Carlo-Collision (PIC-MCC) code designed to model the ion dynamics in the ECRIS plasma. We have choosen to only consider the dynamics of the heavy particles (ions and atoms), calculating the electron density from the requirement of quasi-neutrality and taking the electron temperature as a free parameter. The problem is then greatly simplified in comparison with the situation for the full self-consistent 3-D solution and simulations become computationally affordable. The smallest timescale is defined by the ion movement in the magnetic field, which is in the microsecond range and several orders of magnitude higher than the electron timescale. The largest timescale is determined by the ionization dynamics (around 10 milliseconds).

## CODE DESCRIPTION

The main approximation made in the code is that the influence of a "potential dip"1 on the ion dynamics is negligible. This means that we simulate a quiescent plasma with no gradients of the isotropic electron temperature and without any internal electric fields. The included processes are ion-ion collisions and electron-ion heating in the minimum-B field, as well as the ionization dynamics of ions with different charge states. Neutral particles are tracked in order to take into account the gradients of the gas pressure inside the plasma chamber. Charge-exchange and recombination processes are neglected in our model because at the conditions under exploration their probability is low.

The calculations are done using the geometry and magnetic fields of the KVI-AECRIS, which is described in detail elsewhere[3]. The plasma chamber has an inner diameter of 7.6 cm and a length of 30 cm. The magnetic field has the following properties: the maxima of the solenoidal field are 2.1 and 1.1 T at the injection and extraction sides, respectively, its minimum at the trap centre is 0.36 T and the hexapole field at the radial wall of the source chambers amounts to 0.86 T. The extraction aperture has a diameter of 8 mm. These parameters define the properties of the computational domain of the code.

The solenoid magnetic field of the source is computed by POISSON/SUPERFISH[4], results are stored as a table and imported in the code. Linear interpolation allows calculations of the magnetic field at any requested position. The hexapolar component of the magnetic field is computed by using the analytical expressions for the hexapolar Halbach magnetic structure[5]. External electric fields may be incorporated into the calculations if needed. The computational region is divided into a three-dimensional Cartesian mesh. We use a 38×38×64 mesh in x, y and z directions, respectively, where z is along the source axis.

The ion movement in the magnetic field of the source is tracked using the leap-frog particle mover[6]. The motion of neutral particles is modelled by taking into account scattering and energy losses when hitting the chamber walls.

The collisions between charged particles are computed using the model of Nanbu[7]. In this model a succession of small-angle Coulomb binary collisions is simulated as a single binary collision with a large scattering angle that takes place at each time step. The collision partners are chosen randomly inside a computational cell using the method of Takizuka-Abe[8]. In this way the ion-ion temperature equilibration and ion diffusion processes are taken into account. The ion heating due to electron collisions is calculated by introducing dummy electrons with an isotropic Maxwell-Boltzmann energy distribution. The electron density is defined by requiring charge-neutrality; the electron temperature is a free parameter. The computational time step is chosen to be $10^{-7}$ sec in order to prevent too large scattering angles in the ion-ion collisions.

Knowing the electron density and temperature, the probability for a particle to increase its charge state Q is calculated using the ionization rates from Ref.[9]. These rates are tabulated as a function of the electron temperature for all elements from H to Zn (Z=30). Also, the contributions of excitation-autoionization processes are included in the rates.

When a particle hits the wall of the plasma chamber it is neutralized and returned into the system to be tracked further. Generally, the scattered atom has an energy that depends on the charge of the impinging ion, the plasma potential, the wall material and other factors; at the moment we neglect all these effects and randomly pick the energy from a Maxwell-Boltzmann distribution with a temperature of 1 eV. The angular distribution of the neutralized particle is assumed to be given by a cosine-law[10]. When the neutralized fast atom hits the wall again it looses part of its energy. In the model we set the thermal accommodation coefficient to 0.5. If the energy starts to be comparable to the temperature of the walls, the atom is re-emitted with an energy distribution corresponding to room temperature and in the next collisions the atom does not change its energy anymore. Reaction of the source performance to variations in the described settings will be discussed later.

If particles are lost into the extraction aperture, they are injected back as neutrals at the injection side of the source, with room temperature and with the radius of injection less than 1 cm, imitating the gas injection conditions of the KVI AECRIS. The total number of computational particles is always kept the same ($2\times10^5$); the gas flow into the plasma chamber is automatically adjusted so that the particle flow out of the source through the extraction aperture is always equal to the flow into the source.

The calculations are started with equal numbers of atoms and singly charged particles uniformly distributed over the source volume. This ensemble of particles is then allowed to evolve until a stationary condition is reached, which typically takes around 5 msec. We checked that the initial distribution of the particles does not influence the stationary solution. In the stationary conditions, the extracted ion currents can be calculated from the flux of ions into the extraction aperture, as well as other parameters of the ion population.

The code is written in Compaq Visual Fortran, development is done on a personal computer running Windows XP.

## RESULTS

Calculations have been made to obtain the parameters of ions in the ECRIS plasma for different gas pressures and electron temperatures. We compare the ion outputs when mixing different elements in the discharge, as well as varying the energy and angular distributions of the atoms after ion neutralization on the source walls. Also, changes in the extracted ion currents while varying the source length are discussed briefly.

*Charge-state distributions of ions.*

Numerical solutions with ion currents and other parameters close to the experimentally observed ones can be obtained. In order to achieve this, the gas pressure in the plasma chamber (defined by the particle numerical weight) has to be selected in the range from $10^{-7}$ to $10^{-6}$ mbar and the electron temperature around 1 keV. Both the extracted ion current and the mean charge state increase strongly with increasing gas pressure. This is illustrated in Fig.1, where neon charge-state-distributions (CSD) are shown for different particle weights between $0.75\times10^8$ and $2.25\times10^8$ with a step of $0.25\times10^8$. This corresponds to gas pressures from $4\times10^{-7}$ to $1\times10^{-6}$ mbar, for the 1-keV electron temperature. The arrow in the figure denotes the direction of increasing gas pressure. The experimentally measured distribution of neon ions extracted from the KVI-AECRIS is plotted with open circles. The total extracted ion current varies from 0.2 to 3.8 mA for gas pressures in the above range. The calculations were stopped at a pressure of $10^{-6}$ mbar because the maximum electron density became comparable to the cut-off density for 14.5 GHz microwaves.

No saturation with the gas pressure is observed in the currents, which is in contradiction to experiment. It should be noted, however, that increase in the gas pressure must be accompanied with a corresponding increase in RF power coupled to the plasma to maintain the electron temperature at the same optimal level. In real experimental conditions this is not always possible, the experimentally observed saturation can also be due to onset of plasma instabilities or other factors.

The gas pressure inside the plasma chamber of our source can not be measured directly, but an estimate can be given from the measured value in the vacuum chamber close to the plasma volume; it was around $5\times10^{-7}$ mbar when the experimental spectrum of Fig.1 has been measured. The best correspondence between the simulated and experimental CSD is reached when the particle weights are set to $1.75\times10^8$, corresponding to an average gas pressure of $7.5\times10^{-7}$ mbar, which is close to the experimental values. Given the uncertainties both in the simulations and in measuring the extracted currents, we consider that the agreement between the simulated and measured distributions is good. The same level of agreement is found for all other elements (N, O, Ar) used in the simulations. We take these simulation conditions as the reference ones and characterize the plasma parameters in more detail in the following sections.

*Dependence on the electron temperature*

Dependence of the extracted ion currents on the electron temperature is shown in Fig.2 for the neon gas pressure of 7.6e-7 mbar. A temperature of 1 keV is found to be optimal for maximizing the total extracted ion current and output of the highest charge states for neon. Any increase in the electron temperature above this value results in a decreasing source performance. Generally, high electron temperatures are beneficial for the ion confinement in the plasma, since the ion heating rate is inversely proportional to the square root of the electron temperature. Ionization rates are gradually decreasing after reaching their maxima for an electron temperature around 1 keV. The drop in the ionization rates has a more pronounced effect on the ion production than the decrease in the ion heating rate. If the electron temperature is less then 500 eV, the ion output decreases both due to a drop in the ionization rates and an increase in the heating rate.

*Spatial distributions of the ions*

Large gradients of the ion/electron densities in the radial direction are observed. A projection of the ion trajectories on the y-z plane is shown in Fig.3 – here, a slice is taken in the x- direction with $\Delta x=\pm 1$ mm at x=0. The projection on the x-y plane is shown in Fig. 4. The characteristic six-arm star is due to the hexapole magnetic field. The ion density is strongly peaked on the source axis, and the higher the charge state of the ions, the more compact their spatial distribution is. The maximum electron density reaches a value of $1.1\times10^{12}$ $cm^{-3}$ at the chamber centre, and the plasma diameter is less than 1 cm. The profile of the electron density along the source axis is close to parabolic and the density is steadily decreasing when approaching the extraction and injection sides of the source.

A strong "ion pumping" effect results in variations of the gas density inside the source. At the source axis, the neutral density is around 50% of the average value. This drop in the gas density limits the ion output of the source.

The positions where the ions are hitting the plasma electrode are shown in Fig.5. At the injection side of the plasma chamber, the pattern is basically the same with the three-arm "star" rotated by 180°; it is more compact because of the higher magnetic field. The same tendency as with the spatial distributions of the ions in the plasma is observed – ions with higher charge states hit the plasma electrode at smaller radii. Mean radial sizes of the beam profiles at the plasma electrode are shown in Fig.6 for the reference calculations with neon. The localization of high charge states closer to the source axis is supported by experimental observations of the ion beam normalized emittances as a function of the charge state[11]. A specific feature of the pattern in Fig.5 is that the intense ion flux to the plasma electrode occurs not only at the centre, but also along the axes of symmetry in the star arms, with a relatively broad halo of predominately singly charged ions. This is in good correspondence with the experimental observations of narrow sputtering lines along the star arm axes. Also, the most pronounced sputtering of the electrode is observed at the injection side of the source occurs at the centre, again in the perfect correspondence to the observations.

The mean electron density "seen" by the ions depends on their charge state as shown in Fig.6. Here, the electron density is averaged along the paths of the ions with a specific charge state from the moment of their last ionization till the moment when the ions are lost at the plasma electrode. For the reference neon conditions, the value is around $0.7\times10^{12}$ $cm^{-3}$ for the highest charge state (8+), a factor of 3 higher than for the $Ne^{1+}$ ions. This, again, demonstrates that the highest charge states are localized at the denser parts of the plasma.

The loss profile at the plasma electrode can be compared to the distribution of the magnetic field mirror ratio. The mirror ratio is defined as a ratio of the magnetic field at the given point on the plasma electrode and the minimum value along the magnetic field line connecting to this point, i.e. $R=B_{wall}/B_{min}$. The value (R-1) is plotted in Fig.7 for the magnetic field of the KVI-AECRIS. The "plasma star" at the plate is located where the mirror ratio is highest.

*Ion temperatures*

Temperatures of the ions inside the plasma are calculated to be around 0.25 eV (for neon in the reference conditions). The ions with charge state of 1+ are much colder than the others (0.15 eV), while a weak linear dependence on the charge state is seen for the neon ions with higher charge states. While the velocity distributions for the ions inside the plasma can well be fitted with the Maxwell-Boltzmann expression, the distributions of the lost ions are anisotropic. Significant differences are seen for the velocity distributions along the source axis (along the magnetic field lines) and in the perpendicular directions. Such a distribution is shown in Fig.8 for neon ions. Acceleration of the ions along the source axis is due to the ion pressure gradients. The resulting mean energy of the extracted highly charged ions is a factor of 1.5 higher than the mean energy of ions inside the plasma, i.e. 0.55 and 0.35 eV respectively. As a result of this acceleration the confined ion population is cooled.

*Ion confinement times*

The ion confinement time can be calculated for the different charge states by measuring the intervals between the moment of the singly-charged ion creation and the moment when the ion is lost at the extraction aperture. These first-passage-time distributions (FPTD) are presented in Fig.9 for neon ions for the charge states from 1 to 8+, together with the cumulative distribution for all charge states. The higher the charge state, the longer are their mean times of confinement in the plasma. Reversing the point of view, ions with longer residence times have better chances to be ionized to higher charge states. The tail of the cumulative FPTD can be well fitted with an exponential decay curve; the decay constant is around 0.65 ms for the given conditions.

The mean confinement time can also be defined as the ratio between the total number of ions with given charge-state that stay in the plasma and the current of these ions to the plasma chamber walls. The mean confinement times as a function of the ion charge state are shown in Fig.10 for neon ions. It is close to linear for the lower charge states and then saturates at a level of 0.6 ms. Good agreement is found between these data and measurements[12], where the ion confinement times were estimated for an argon plasma from the intensity of the characteristic X-ray emission. It should be noted that the residence times of Figs.9 and 10 reflect both the losses of the ions to the chamber walls and losses due to ionization into the higher charge states. For the highest charge states, the loss branch to the chamber walls is dominant because of the low ionization probability.

The time distributions shown in Fig.9 can also be compared with the measurements of the charge-breeding of ions after pulsed injection of atoms into the ECR plasma by the laser ablation technique[13]. The same tendencies are observed for the times needed to produce the highest charge-state ions, as well as for the estimations of the effective electron density and temperature.

The ion confinement time in the so-called gas-dynamics mode, that is in the limit of a zero “potential dip”, is given[14] by $\tau_c = RL/v_i$, where R is the magnetic mirror ratio, L is the characteristic length of the plasma, $v_i$ is thermal velocity of ions. The mirror ratio along the source axis for the extraction side of our source is equal to 3, L is 15 cm and the mean velocity is $1.6\times10^5$ cm/sec for the $Ne^{6+}$ ions in the reference condition, corresponding to the thermal velocity of $1.0\times10^5$ cm/sec; the resulting estimate for the confinement time is 0.45 ms, in reasonable agreement with the data. The gas-dynamics confinement defines the scale of plasma production in our simulations.

*Gas-mixing and isotope anomaly*

After characterizing the ion output from the ECRIS plasma in the reference conditions, we present some results obtained by varying specific source parameters. First, it was checked whether the isotope anomaly [2] can be reproduced by the model. This was done by running simulations starting with an equal mixture of $^{20}$Ne and $^{22}$Ne isotopes and keeping the isotope fluxes into the source equal to each other during the run.

A pronounced effect is observed for the ratio of extracted currents for both isotopes as a function of the ion charge state, see Fig.11. The current of $^{22}Ne^{9+}$ is 15% higher than the $^{20}Ne^{9+}$ current, while the currents of the low charge states for the heavier isotope are 5% smaller. The same effect for neon isotopes is also experimentally observed at our source.

Differences in the confinement times for the isotopes are observed: the slopes of the FPTD tail as shown in Fig.9 are 0.67 ms for $^{20}$Ne and 0.69 ms for $^{22}$Ne ions in the given plasma conditions. The ratio of these values scales approximately as the square root of the mass ratio and corresponds to the ratio between the mean velocities of the ions. Even very small differences in the decay constant causes pronounced relative variations at large time scales.

Also, quite different spatial distributions are observed for the isotopes – beam profiles of the heavier $^{22}$Ne ions at the plasma electrode are systematically more compact then for the lighter isotope. This can be understood in the terms of the slower outward diffusion of the heavier particles in the radial direction. The reason for this is that the diffusion rate is determined by the ion thermal velocities and not by the Larmor radius when the ion-ion collision rate is much higher than the cyclotron frequency. Taking into account the large radial gradients of the plasma density, the heavier isotope stays in the regions with larger mean electron density (up to 2%) when it diffuses through the plasma. Combined together, these two factors lead to more effective production and extraction of the higher charge states for the heavier isotopes.

The gas-mixing effect [2] has also been studied for argon and neon in simulations where these gases are mixed with helium/oxygen in different proportions. The results of these simulations were compared with those of simulations of Ar/Ne plasmas having the same particle numerical weights and approximately the same total gas pressure. A positive gas mixing effect was not observed. This may be connected to the fact that our simulations do not reproduce the experimentally observed decrease in source performance at higher gas pressures. Ion temperatures are not significantly affected by addition of the lighter gases into the discharge, indicating small changes in the ion cooling rate. We conclude therefore that the physical mechanisms for the isotope anomaly and for the gas-mixing effect are different.

*Wall coating effect*

It is known that performance of ECR ion sources is strongly affected by the wall conditions in the source. In particular, the extracted currents of highly charged ions are increased noticeably when the walls are covered with thin films of oxides. This so-called wall coating effect [2] is usually explained in terms of the higher secondary electron emission from oxidized surfaces, which results in a lower plasma potential.

An alternative (complimentary) approach is connected to the influence of the wall conditions on the energy and angular distributions of the fast back-scattered atoms created after wall neutralization of ions. Indeed, before the ions are lost at the walls they acquire in the plasma sheath a relatively high energy, proportional to the plasma potential (typical experimental values of the plasma potential are a few tens of Volts) and the charge state of the ion. The direction of this acceleration is along the surface normal. For ion energies in the range of a few tens of eV, the scattered atoms retain up to 5% of their primary energy depending on the wall material, surface conditions, ion mass and other factors[15]. This results in the creation of fluxes of relatively energetic atoms, which in turn leads to extra heating of the ion population after ionization of these atoms.

Reflection of atoms from the surface can be described as mainly a combination of diffusive and specular scattering. The diffusive reflection is considered to result from multiple scattering from a rough surface or from the layers close beneath the surface. The angular distribution of the scattered atoms in this case is given by Lambert's cosine law. The specular reflection assumes single collisions between the impinging ion and solid atoms. In the situation where the main acceleration of the ions occurs towards the surface along the surface normal, the specular reflected atoms will be directed in the opposite direction.

We studied the influence of fast atoms in an approximation where they have a Maxwell-Boltzmann energy distribution with a certain temperature, while neglecting the variations in the temperature for different charge states of the primary ions. Plasma parameters are then calculated for temperatures varying from 0.025 eV (wall temperature) to 10 eV.

Simulations show that if the scattering angles of the atoms are defined according to the "cosine-law", there is quite a small influence of the atom excess energies on the plasma. The atoms are scattered and cooled at the walls many times before they are ionized and trapped in the plasma and then transfer their energy to the ion population.

The situation is different when we assume instead that all neutralized ions are back reflected at normal angles to the wall. These atoms are then effectively ionized before their thermalization in collisions with the walls. The resulting ion heating leads to a loss of confinement. It is observed that the ion output is decreased drastically if the energy of the scattered atoms is above of a few eV.

It is rather difficult to estimate realistic distributions of atom scattering angles and energies inside the source. The general conclusion from our simulations is that the diffusive scattering from the walls is more preferable for optimizing the source performance. The way to control the fast atom distributions is to use oxidized surfaces. These are known to have higher energy absorption coefficients and higher scattering ability, which results in a decrease of the gas temperature inside the source.

*Source length scaling*

Preliminary studies have been performed on the effect of increasing the source dimensions. This was done by doubling the plasma chamber length. We have used the same solenoidal magnetic field as in the reference conditions, but have decreased the radial component by a factor of two to respect Maxwell's equations. The hexapolar field was not changed. To keep the gas pressure at the same level as before, the numerical weight of the particles was doubled. The charge-state-distribution of the extracted neon ions is plotted in Fig.13 together with the CSD for the original chamber length. The increase in the currents of the highest charge states of the neon ions is remarkable. Simultaneously, longer ion confinement times are observed, confirming that these times are proportional to the plasma length. We conclude that for producing the highest charge states, the long sources are preferable.

## CONCLUSION

The results of our simulations clearly show the importance of taking into account the diffusion processes in the ECRIS plasma when modelling the ion dynamics. The extracted ion currents and shape of the plasma are well reproduced by the code assuming the gas-dynamics mode for the ion confinement. This does not exclude the existence of electric fields that confine the ions in the plasma, especially at low gas pressure, but indicates that such effects are probably of minor importance. It is shown that the long-standing problem of the isotope anomaly is connected to the ion diffusion dynamics in the plasma. The importance of taking into consideration the gas pressure gradients inside the plasma and parameters of the fast back-scattered atoms is confirmed by the code.

The model can be used for studies of different effects in ECRIS plasmas, such as scaling of the source dimensions and of the magnetic field structure. Application of the model to simulations of charge breeding in on-line ECR ion sources is feasible and will require only small modifications of the code.

## FIGURE CAPTIONS

Fig.1 Charge-state-distributions of the extracted Ne ions as a function of the gas pressure. Arrow shows the direction of increasing gas pressure. Curves correspond to the gas pressures (3.8, 4.8, 5.8, 6.8, 7.6, 8.5, 9.3)$\times 10^{-7}$ mbar. Open circles represent experimentally measured extracted ion currents from the KVI-AECRIS.

Fig.2 Charge state distributions of the extracted neon ions for different electron temperatures. Gas pressure is $7.6\times 10^{-7}$ mbar.

Fig.3 The y-z projection of ion trajectories with a slice of x=0±1 mm.

Fig.4 The x-y projection of the ion trajectories.

Fig.5 Spatial distribution of the lost neon ions at the plasma electrode.

Fig.6 Mean radius of the ion beam profile at the plasma electrode (squares) and mean electron density (circles) seen by the ions for the different charge states of neon.

Fig.7 Mirror ratio of the magnetic field ($B_{wall}/B_{min}$-1) at the plasma electrode for KVI-AECRIS.

Fig.8 Velocity distributions in y and z-directions of the neon ions that are lost at the plasma electrode.

Fig.9 Distribution of residence times inside the plasma for the different charge states of neon from 1+ to 8+.

Fig.10 Dependence of the mean confinement time on the ion charge state for neon.

Fig.11 Ratio between the extracted currents of $^{22}Ne$ and $^{20}Ne$ ions as a function of the charge state.

Fig.12 Mean radius of the ion beam at the plasma electrode (squares) and mean electron density (circles) seen by the ions for different charge states of neon. Open symbols are for the $^{22}Ne$ ions, solid ones are for $^{20}Ne$.

Fig.13 Charge state distributions of the extracted neon ions for the basic conditions (green columns) and for the doubled plasma chamber length (red columns).

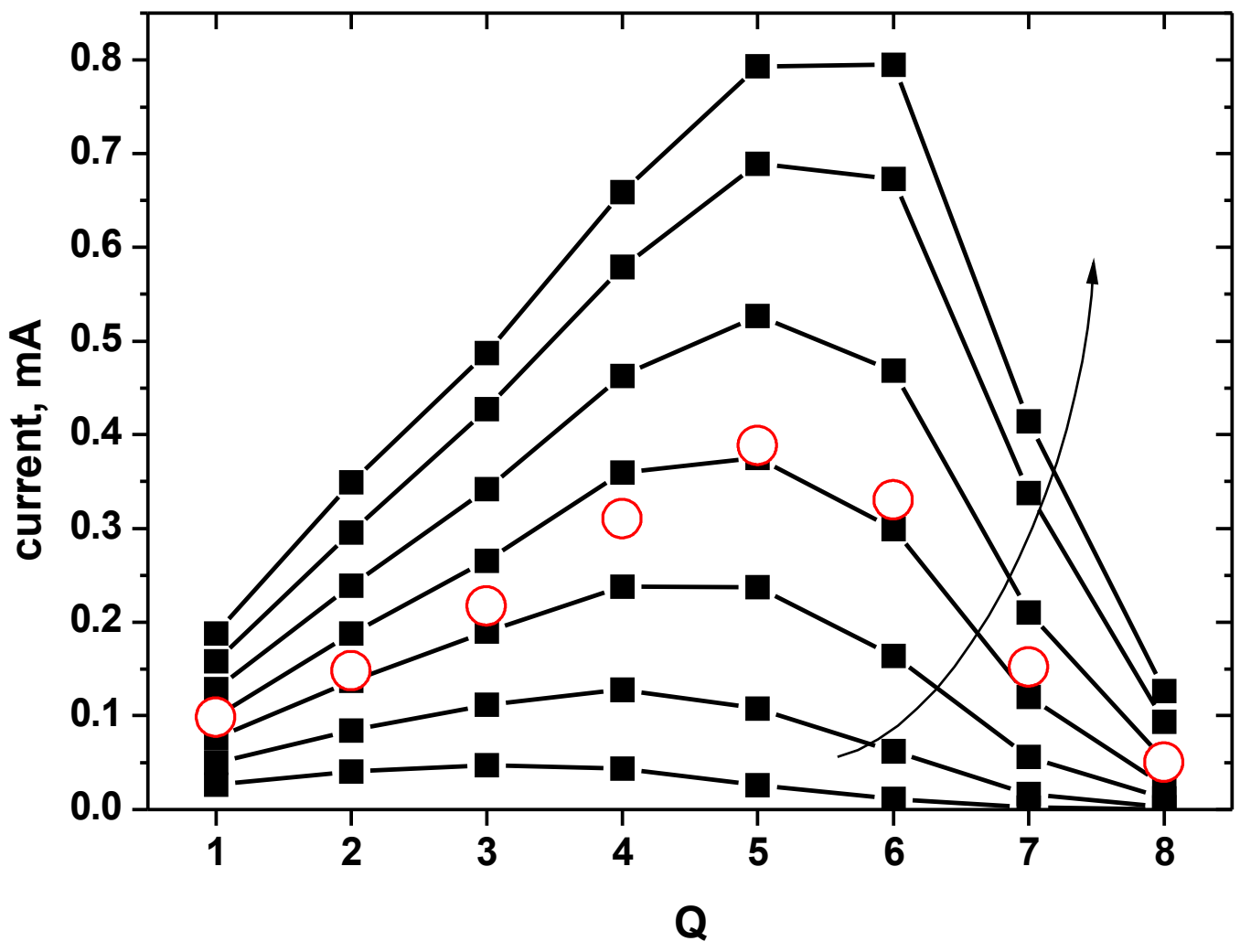

V. Mironov, J.P.M.Beijers, Three-dimensional simulations…

Fig.1 Charge-state-distributions of the extracted Ne ions as a function of the gas pressure. Arrow shows the direction of increasing gas pressure. Curves correspond to the gas pressures (3.8, 4.8, 5.8, 6.8, 7.6, 8.5, 9.3)$\times 10^{-7}$ mbar. Open circles represent experimentally measured extracted ion currents from the KVI-AECRIS.

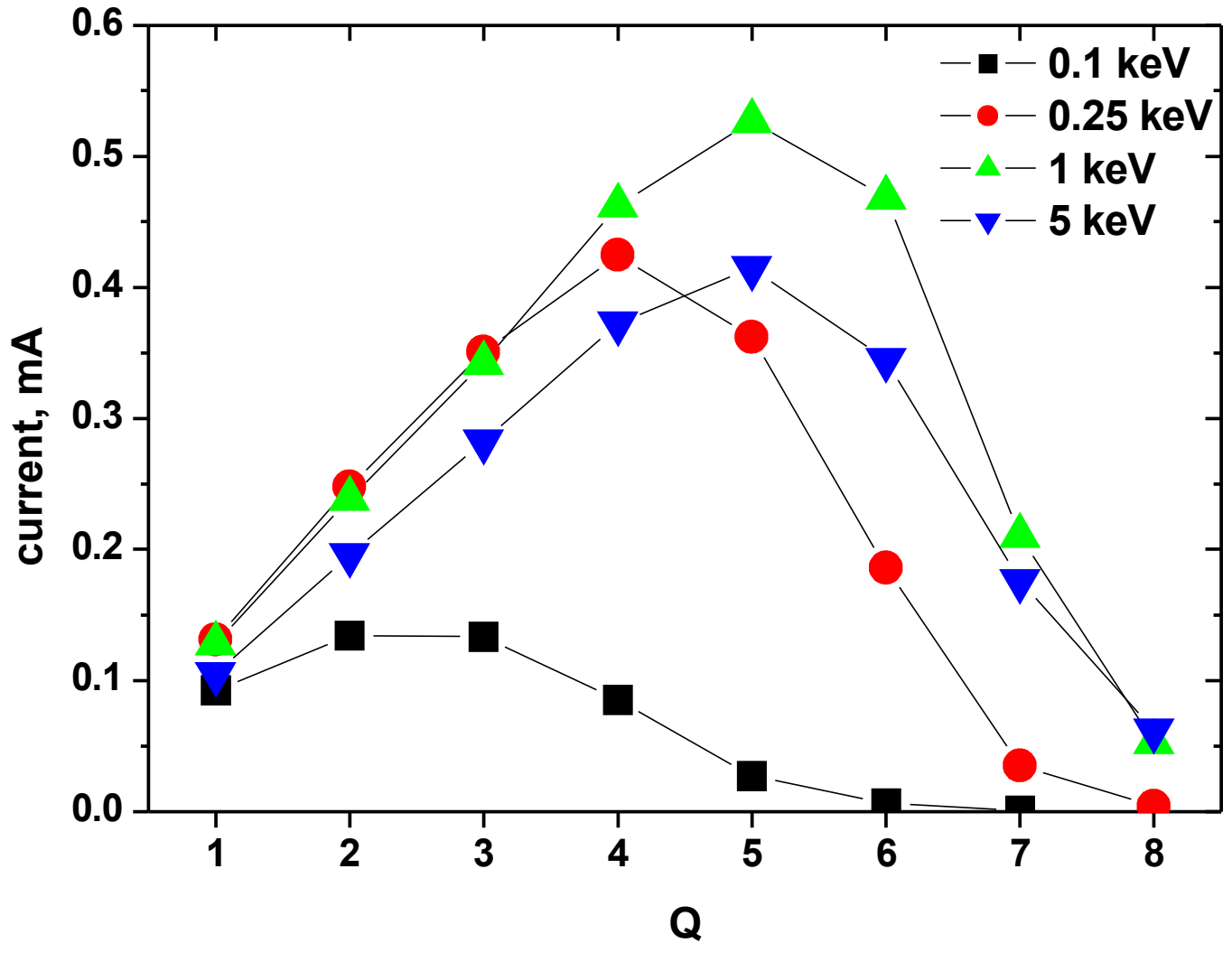

V. Mironov, J.P.M.Beijers, Three-dimensional simulations…

Fig.2 Charge state distributions of the extracted neon ions for different electron temperatures. Gas pressure is $7.6\times10^{-7}$ mbar.

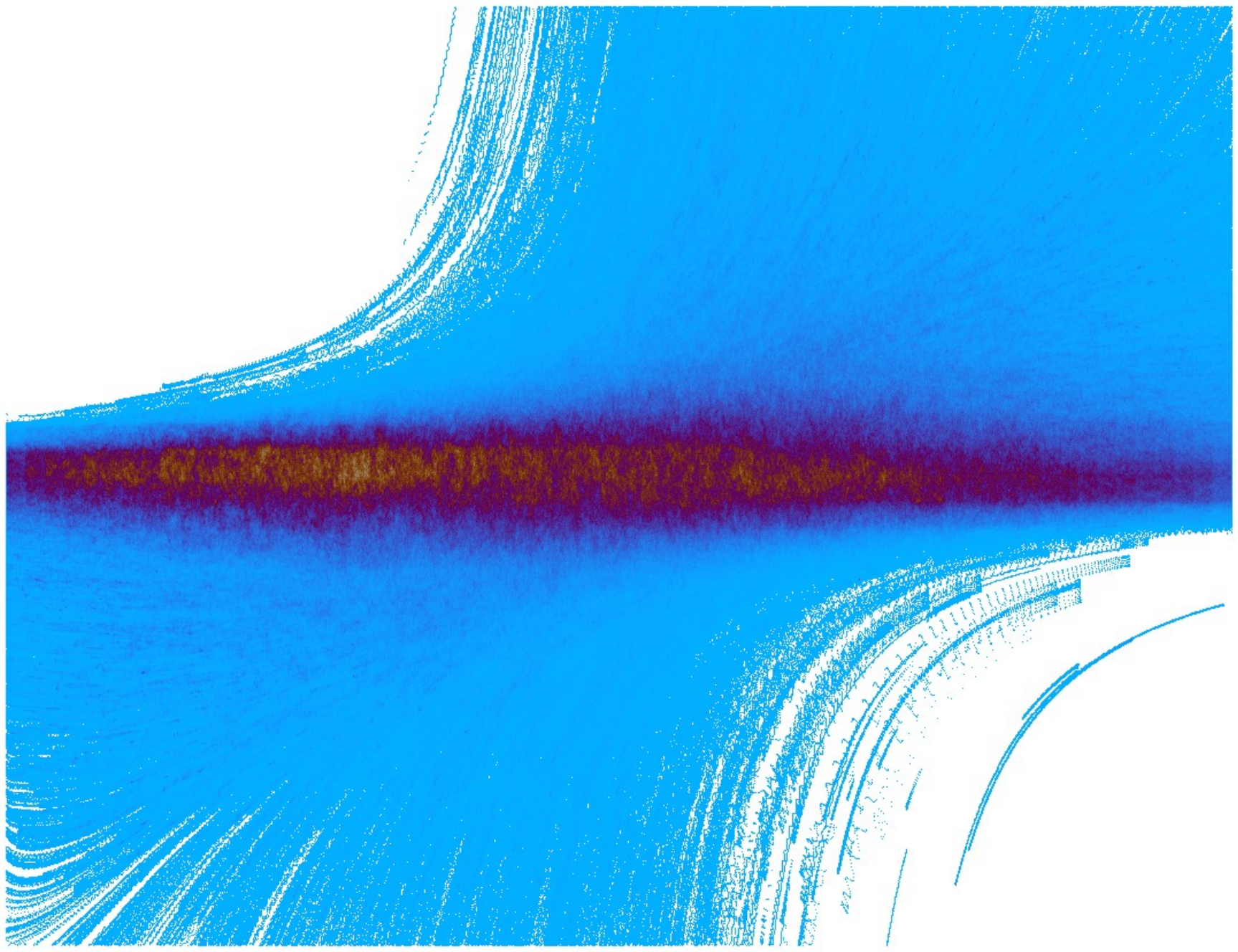

V. Mironov, J.P.M.Beijers, Three-dimensional simulations…

Fig.3 The y-z projection of ion trajectories with a slice of x=0±1 mm.

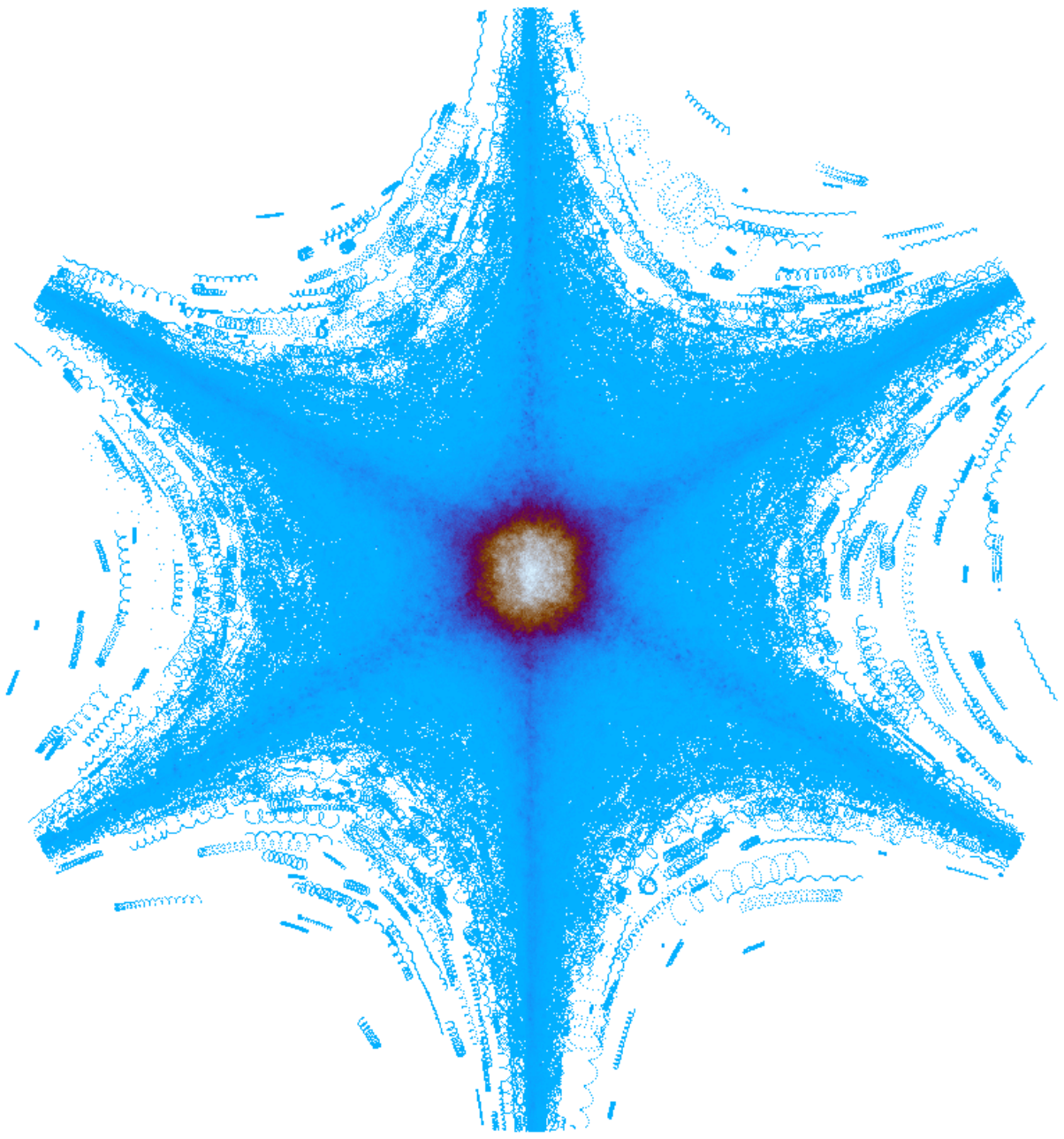

V. Mironov, J.P.M.Beijers, Three-dimensional simulations…

Fig.4 The x-y projection of the ion trajectories.

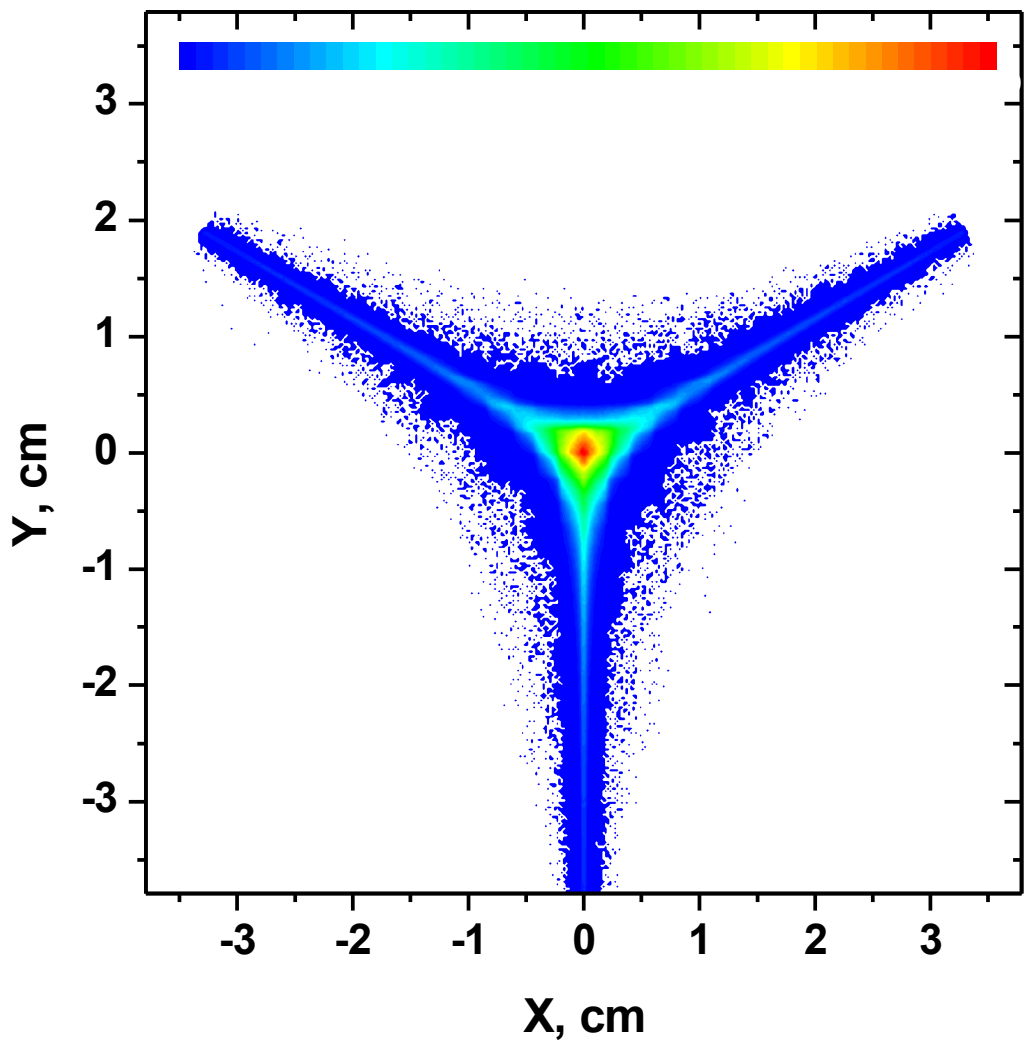

V. Mironov, J.P.M.Beijers, Three-dimensional simulations…

Fig.5 Spatial distribution of the lost neon ions at the plasma electrode.

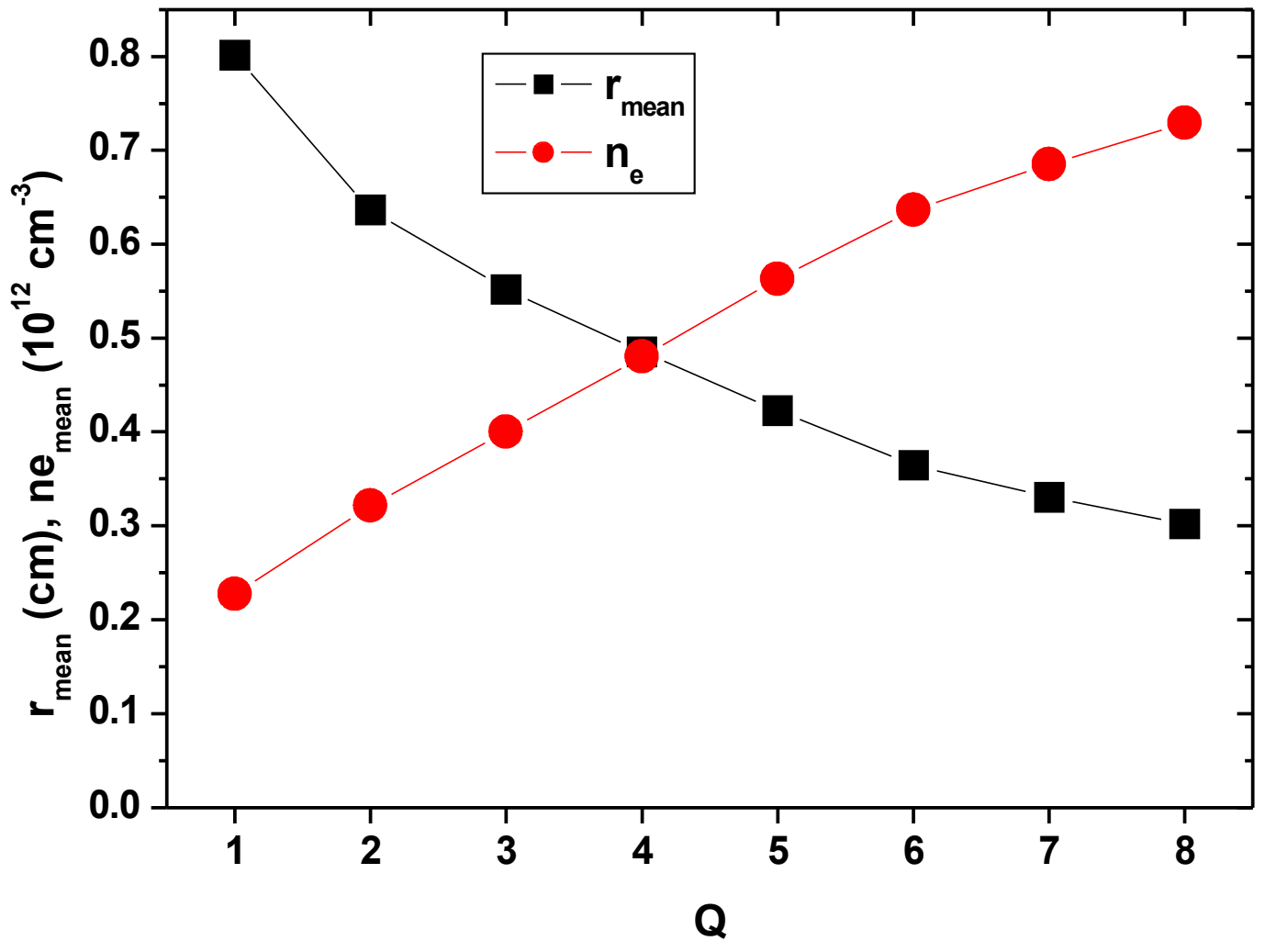

V. Mironov, J.P.M.Beijers, Three-dimensional simulations…

Fig.6 Mean radius of the ion beam profile at the plasma electrode (squares) and mean electron density (circles) seen by the ions for the different charge states of neon.

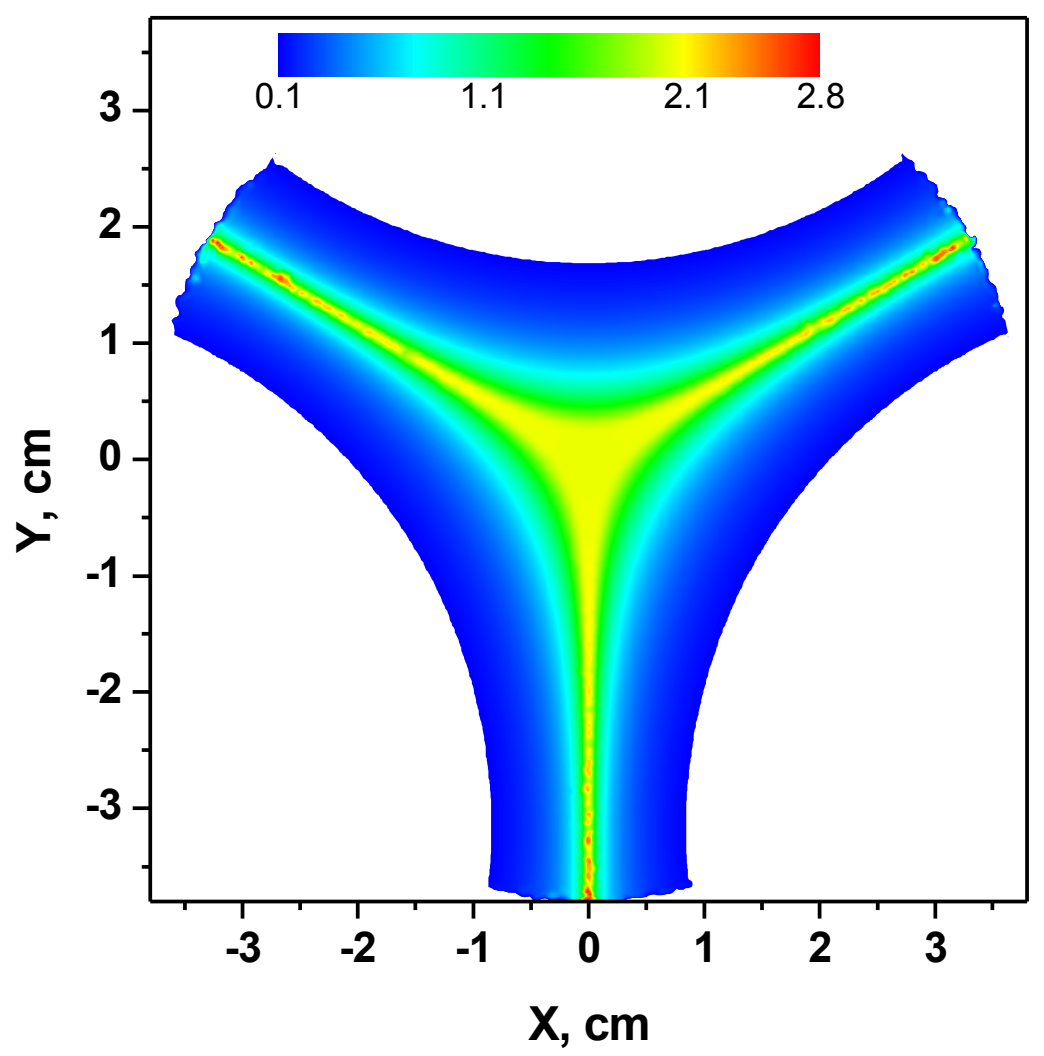

V. Mironov, J.P.M.Beijers, Three-dimensional simulations…

Fig.7 Mirror ratio of the magnetic field ($B_{wall}/B_{min}$-1) at the plasma electrode for KVI-AECRIS.

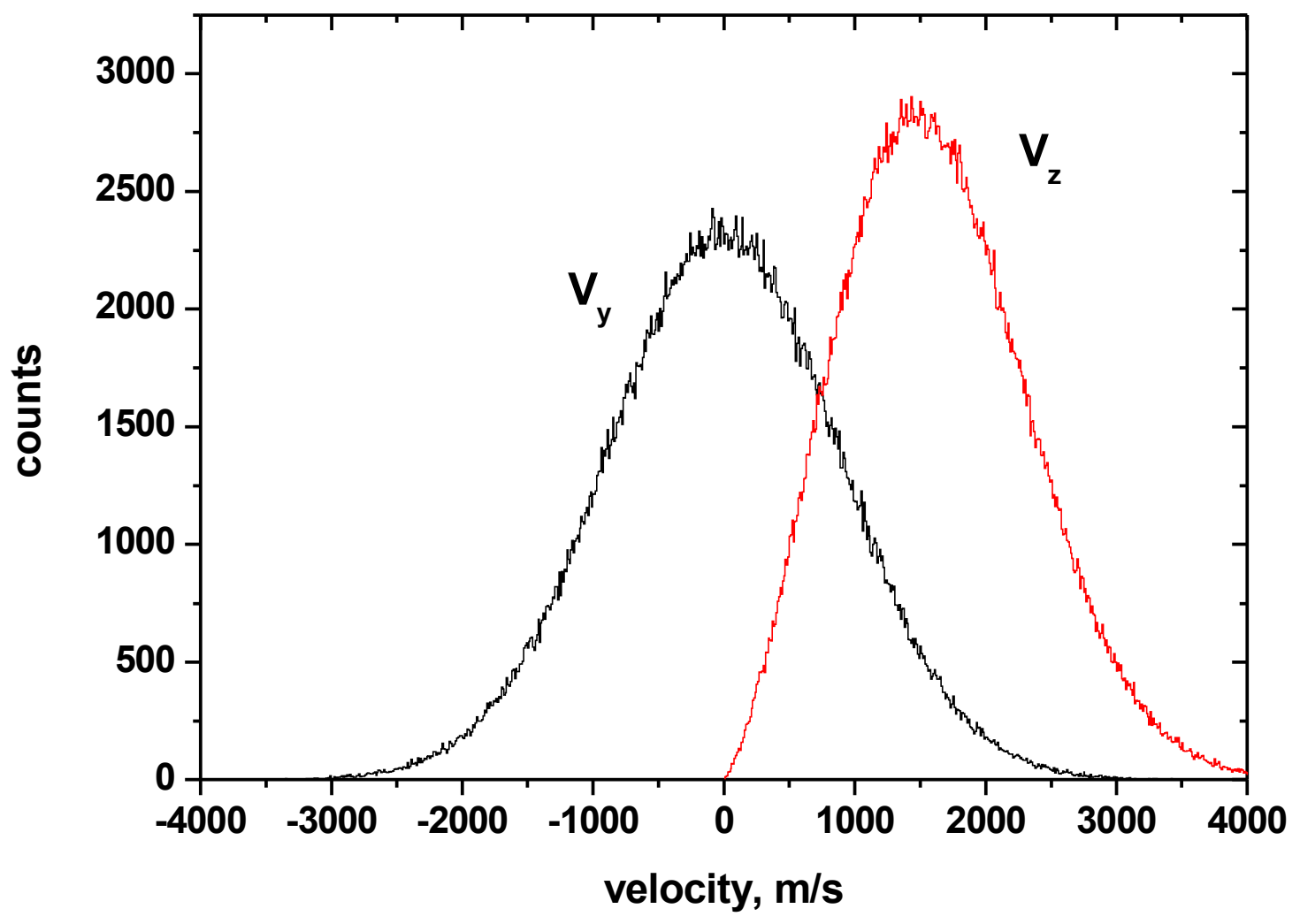

V. Mironov, J.P.M.Beijers, Three-dimensional simulations…

Fig.8 Velocity distributions in y and z-directions of the neon ions that are lost at the plasma electrode.

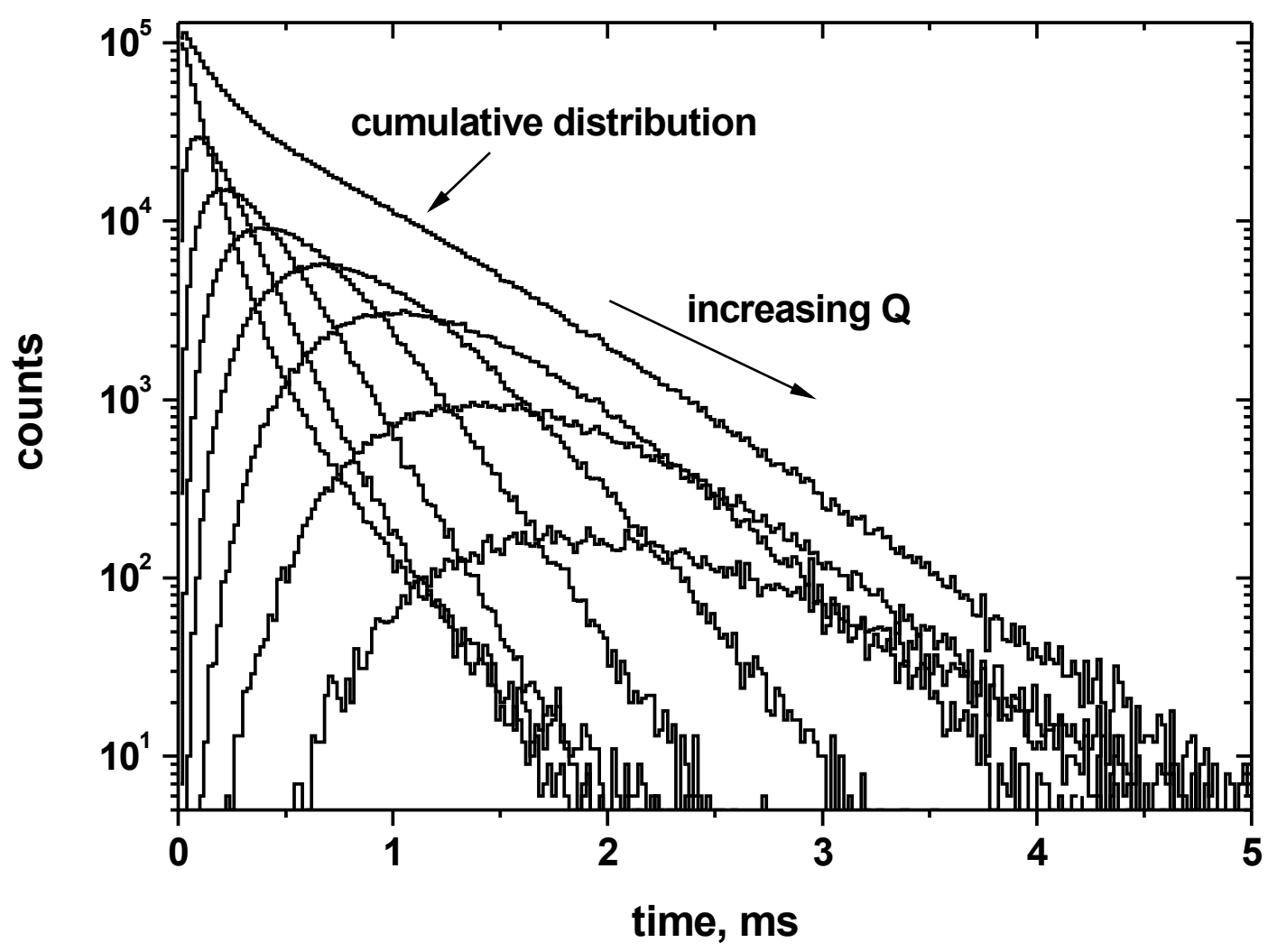

V. Mironov, J.P.M.Beijers, Three-dimensional simulations…

Fig.9 Distribution of residence times inside the plasma for the different charge states of neon from 1+ to 8+.

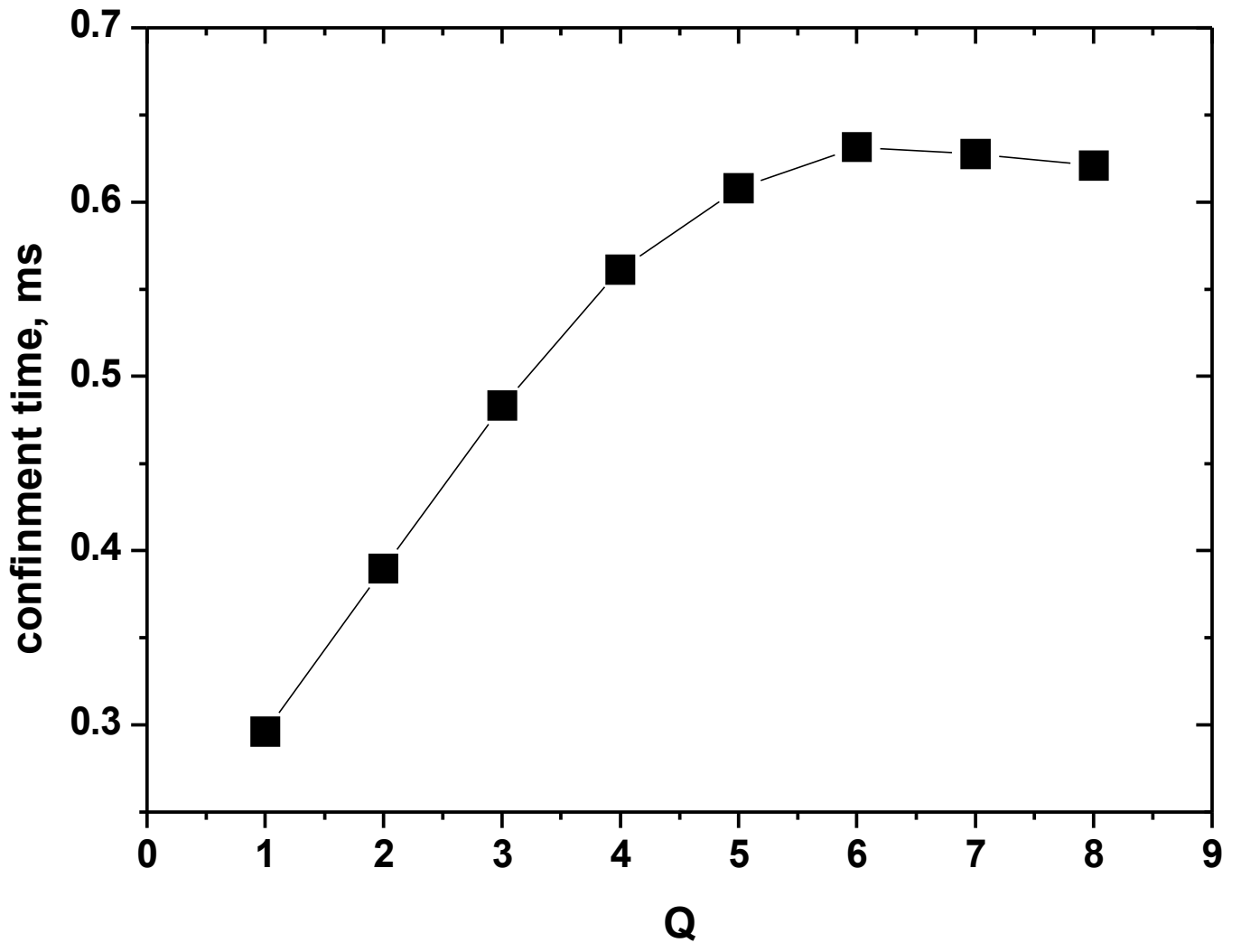

V. Mironov, J.P.M.Beijers, Three-dimensional simulations…

Fig.10 Dependence of the mean confinement time on the ion charge state for neon.

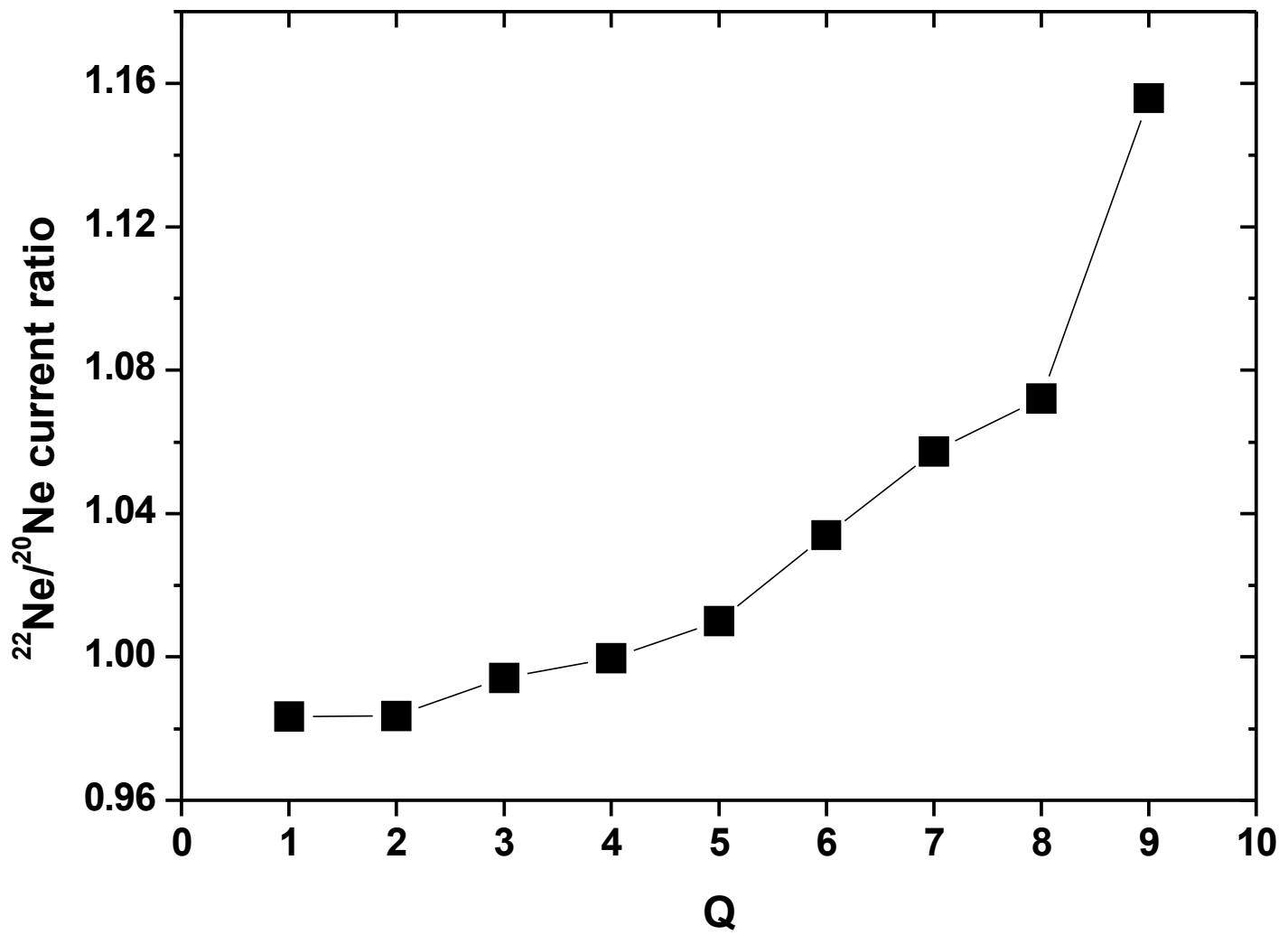

V. Mironov, J.P.M.Beijers, Three-dimensional simulations…

Fig.11 Ratio between the extracted currents of $^{22}Ne$ and $^{20}Ne$ ions as a function of the charge state.

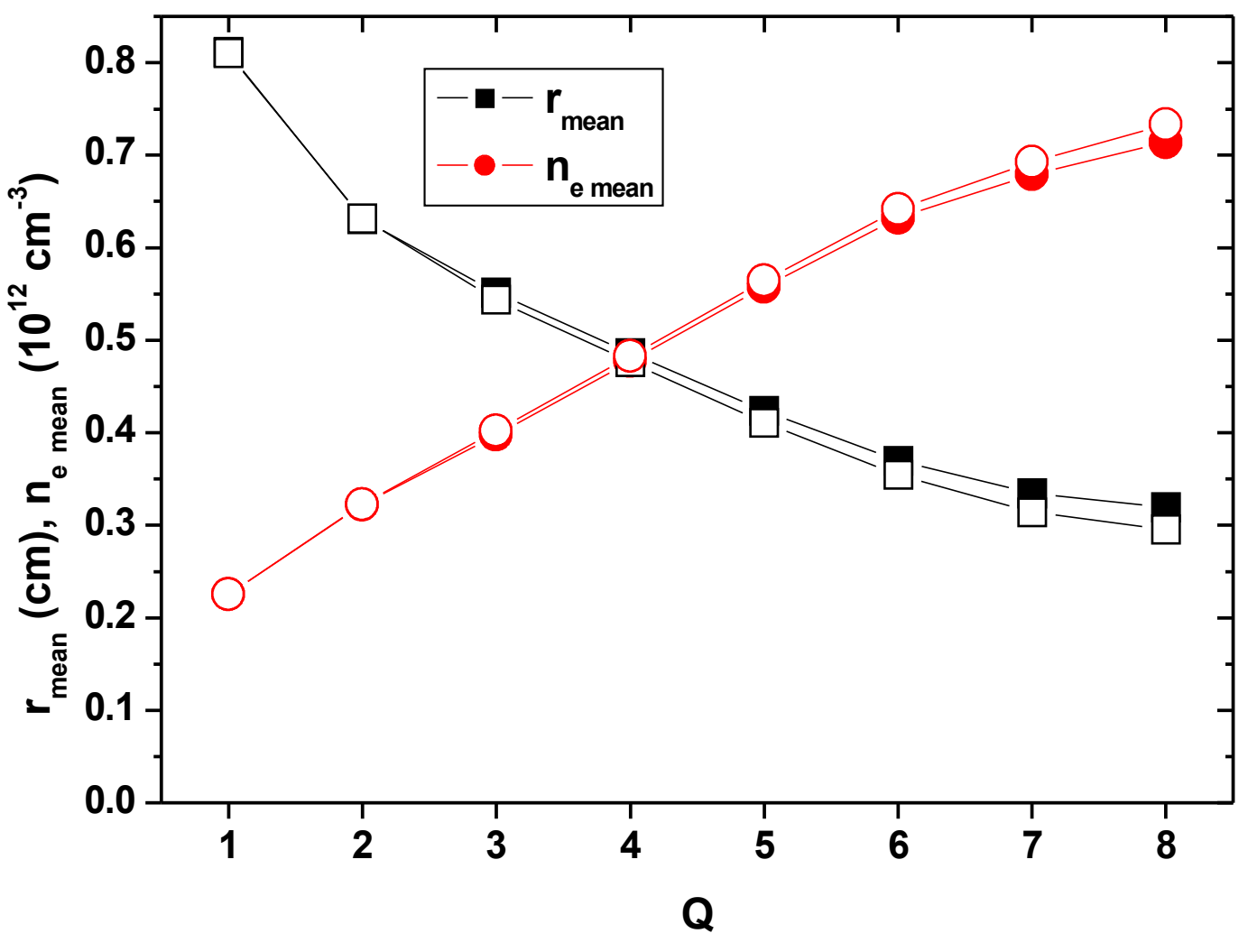

V. Mironov, J.P.M.Beijers, Three-dimensional simulations…

Fig.12 Mean radius of the ion beam at the plasma electrode (squares) and mean electron density (circles) seen by the ions for different charge states of neon. Open symbols are for the $^{22}Ne$ ions, solid ones are for $^{20}Ne$.

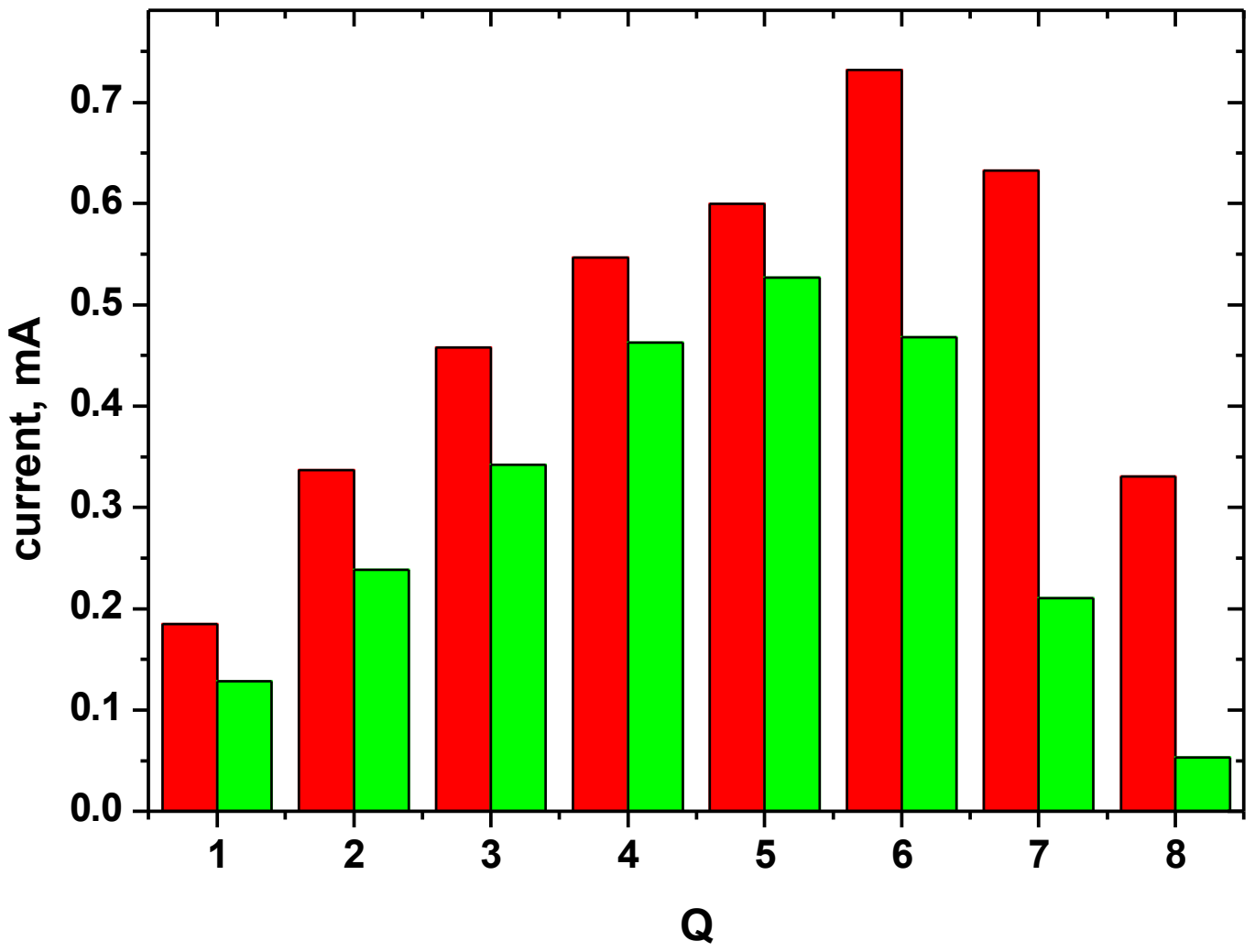

V. Mironov, J.P.M.Beijers, Three-dimensional simulations…

Fig.13 Charge state distributions of the extracted neon ions for the reference conditions (green columns) and for the doubled plasma chamber length (red columns).

**REFERENCES**

[1] R.Geller, Electron Cyclotron Resonance Ion Sources and ECR Plasma (Bristol: Institute of Physics), 1996.
[2] A.G.Drentje, Rev. Sci. Instrum. **74**, 2631 (2003) and references therein.
[3] H.R.Kremers, J.P.M.Beijers and S.Brandenburg, Rev. Sci. Instrum. **77**, 03A311 (2006).
[4] http://laacg1.lanl.gov/laacg
[5] K.Halbach, Nucl.Instrum.Methods **169**, 1 (1980).
[6] J.P. Verboncoeur, A.B. Langdon and N.T. Gladd, Comp. Phys. Comm. **87**, 199 (1995).
[7] K.Nanbu, Phys.Rev.E. **55**, 4642 (1997).
[8] T.Takizuka and H.Abe, J.Comp.Phys. **25**, 205 (1977).
[9] P.Mazzotta P, G.Mazzitelli, S.Colfrancesco, N.Vittorio, Astron. Astrophys., Suppl. Ser., **133**, 403 (1998).
[10] J.Greenwood, Vacuum **67**, 217 (2002).
[11] D.Wutte, M.A.Leitner and C.M.Lyneis, Phys.Scr. **T92**, 247 (2001).
[12] G.Douysset, H.Khodja, A.Girard and J.P.Briand, Phys.Rev.E **61**, 3015 (2000).
[13] V.Mironov, S.Runkel, K.E.Stiebing, Rev.Sci.Instrum. **72**, 2271 (2001).
[14] T.Rognlien and T.Cutler, Nucl.Fusion **20**,1003 (1980).
[15] H.F.Winters, H.Coufal, C.T.Rettner and D.S.Bethune, Phys.Rev.B **41**, 6240 (1990).